\documentclass[prc,aps,eqsecnum,twocolumn,floatfix,nofootinbib]{revtex4-1}
\usepackage{ulem}  
\usepackage{dcolumn}
\usepackage{bm}
\usepackage{color}
\usepackage{amssymb}
\usepackage{amsmath}
\usepackage{graphicx}
\usepackage{amsfonts}
\usepackage{slashed}
\usepackage{pstricks}
\usepackage{float}
\usepackage{hyperref}
\usepackage{array}
\usepackage{epsf}
\usepackage{soul}
\usepackage{siunitx}
\usepackage{multirow}
\usepackage{subfigure}

\newcommand{\m}{\phantom{-}}

\allowdisplaybreaks
\begin{document}
\title{Bayesian analysis of muon capture on deuteron in chiral effective field theory}
\author{A.\ Gnech$^{\rm a,b}$, L.E. Marcucci$^{\rm c,d}$ and M. Viviani$^d$}
\affiliation{
  $^{\rm a}$\mbox{European Center for Theoretical Studies in Nuclear Physics and Related Areas (ECT$^*$)}
\mbox{ and Fondazione Bruno Kessler
  Strada delle Tabarelle 286, I-38123 Villazzano (TN), Italy}\\
  $^{\rm b}$\mbox{INFN-TIFPA Trento Institute of Fundamental Physics and Applications, Via Sommarive, 14, I-38123 Trento, Italy}
  $^{\rm c}$\mbox{Dipartimento di Fisica “E. Fermi”, Universit\`a di Pisa, Pisa I-56127, Italy}
  $^{\rm d}$\mbox{Istituto Nazionale di Fisica Nucleare, Sezione di Pisa, Pisa I-56127, Italy}
}
\date{\today}
\begin{abstract}
  We compute the muon capture on deuteron in the doublet hyperfine state for
  a variety of nuclear interactions and consistent nuclear currents. 
  Our analysis includes a detailed examination
  of the theoretical uncertainties coming from different
  sources: the single-nucleon axial form factor,
  the truncation of the interaction and current
  chiral expansion, and the model dependence.
  Moreover, we study the impact of the use of different power counting scheme for the electroweak currents on the truncation error.
  To estimate the truncation error of the chiral expansion of interactions
  and currents we use the most modern techniques
  based on Bayesian analysis. This method enables us to give a clear statistical interpretation of the computed theoretical uncertainties. Finally, 
  we provide the differential capture rate
  as function of the kinetic energy of the outgoing
  neutron which may be  measured in future experiments.
  Our recommended theoretical value for the total  doublet capture rate is
  $\Gamma_{\rm th}=395\pm 10$ s$^{-1}$ ($68\%$ confidence level).
  We calculated also the capture rate in the quartet hyperfine state, which turns out to be in the range $[13.3-13.8]$ s$^{-1}$ depending on the adopted nuclear interaction.
 \end{abstract}
 
\index{}\maketitle
\section{Introduction}
\label{sec:intro}
The muon capture on deuteron, i.e. the process
\begin{equation}\label{eq:reaction}
  \mu^-+d\rightarrow n+n+\nu_\mu
\end{equation}
is one of the electroweak processes
that is accessible experimentally and at the same time can be
computed using the most modern theoretical nuclear physics techniques combining
chiral effective field theory ($\chi$EFT) with {\it ab-initio} numerical methods.
This makes the reaction in Eq.~(\ref{eq:reaction}) the ideal system
for testing $\chi$EFT interactions and electroweak
currents.

The experimental results for the total capture rate in the initial doublet hyperfine state $\Gamma$~\cite{Wang1965,Bertin1973,Bardin1986,Cargnelli1989}
are relatively old and  have
large error-bars that make them compatible
within each other but 
hard to use for precision tests of the theory. To improve the precision of experimental measurements,
an on-going experiment at the
Paul Scherrer Institute, in Switzerland, performed by the MuSun Collaboration, aims
to reduce the uncertainty at the order of $\approx1.5\%$~\cite{Kammel2021}. If such precision is  achieved, it will enable more stringent test of the $\chi$EFT predictions.

On the theory side,
several calculations have been carried out.
A review of the theoretical results
from up to about ten years ago can be found in Ref.~\cite{Marcucci2012review}. 
Some more recent calculations, but not yet fully within the
$\chi$EFT framework, have been performed also in Refs.~\cite{Marcucci2011,Golak2014}. The first steps within
the ``hybrid" $\chi$EFT approach, where phenomenological potentials
are used together with $\chi$EFT nuclear currents, have been performed in Refs.~\cite{Marcucci2011,Ando2002}. On the other hand, the first fully consistent $\chi$EFT calculations are those of Refs.~\cite{Marcucci2012chiral,Adam2011}. To be
remarked that the results of Ref.~\cite{Marcucci2012chiral}, as well as those of Ref.~\cite{Adam2011}, were affected by a widely-spread error in the relation between the low-energy constant (LEC) entering the three-nucleon interaction and that one entering the axial current. In Ref.~\cite{Marcucci2019} such error was spotted and the results of Ref.~\cite{Marcucci2012chiral} were corrected. In the last years, a fully consistent $\chi$EFT calculation, not affected by the above mentioned error, has been performed in Ref.~\cite{Acharya2018}. This work addressed also 
the impact of the experimental uncertainty
 on the single-nucleon axial form factor~\cite{Hill2018}. To be noticed that the work in Ref.~\cite{Acharya2018} retained only the $^1S_0$ channel while in that  
of Bonilla {\it et al.} in Ref.~\cite{Bonilla2023}  a complete partial wave expansion was performed together with an analysis of the theoretical uncertainties to study the capture rate in both the $f=1/2$ (doublet) and $f=3/2$ (quartet) hyperfine states.

This work moves on a parallel line, starting from our recent
paper where we considered only the $^1S_0$ neutron-neutron channel
~\cite{Ceccarelli2023}.
We complete the calculation of Ref.~\cite{Ceccarelli2023}
adding the missing channels with a dual purpose.
The first is to have the most robust theoretical error estimate
based on the Bayesian analysis of the truncation errors of the chiral
current and interaction expansions, on the model dependence, and on the propagation of the
uncertainties related to the LECs appearing in the currents.  Moreover, we carry on also an analysis of the impact of the use of different power counting for the electroweak currents on the truncation errors.  For the Bayesian analysis we use the approach
introduced in Ref.~\cite{Melendez2019} and
already employed in several works to give reliable estimates of truncation errors in chiral effective field theory (see  Ref.~\cite{buqeye} for a complete list of works). 
The second purpose is to compute the spectra of the outgoing neutrons as function of the
kinetic energy of the neutron. Beyond the technical details,
this kind of spectrum can possibly be measured in MuSun experiment~\cite{Andreev2010} and can be
useful for the simulation of the experimental apparatus, and consequently for data analysis.

The paper is organized as follow. In the next section
we will introduce the theoretical formalism giving the explicit
expression for the differential capture rate as function of the kinetic energy of
the neutron. In Section~\ref{sec:models} we present the nuclear interaction and current
models used in this work. Section~\ref{sec:errors} is dedicated to a detailed
examination of the theoretical uncertainties. In Section~\ref{sec:discussion},
we  discuss our results comparing them with the recent literature. Finally, in Section~\ref{sec:musun} we consider
the impact of our results on the analysis of the future data of
the MuSun experiment.

\section{Muon capture on deuteron fundamentals}
\label{sec:xsec}
In the past literature the muon capture differential
capture rate was computed versus the relative momenta of the
two emitted neutrons. This has the advantage to reduce the numerical effort
needed in the calculation.
On the other hand, this is not what can be measured experimentally.
In this work we  consider the differential capture rate as function
of the kinetic energy of one of the emitted neutrons (i.e. $E_1'=E_1-m_n$), 
which is the quantity that can potentially be measured in an experiment 
using a neutron detector.
Let us begin with the Fermi Golden Rule, that reads
\begin{equation}\label{eq:fermigd}
  d\Gamma=(2\pi)\delta(E_1+E_2+k_\nu-m_\mu-m_d)\overline{|T_{fi}|^2}
  \frac{d{\bf p}_1}{(2\pi)^3}\frac{d{\bf k}_\nu}{(2\pi)^3}\,,
\end{equation}
where $E_{1}(E_{2})$ is the energy of the first (second) outgoing neutron,
$k_\nu$ the energy of the emitted neutrino,  ${\bf p}_1 ({\bf p}_2)$ the momenta of 
the first (second) outgoing neutron and ${\bf k}_\nu$ the momentum of the outgoing
neutrino. Note that the phase space of the second emitted
neutron has been eliminated using the conservation of the momenta.
The transition amplitude is written as in Ref.~\cite{Marcucci2011}
\begin{equation}
\overline{|T_{fi}|^2} = \frac{1}{2f+1}\sum_{s_1 s_2 h_\nu}\sum_{f_z}
|T_{fi}(f,f_z;s_1,s_2, h_\nu)|^2 \ ,
\label{eq:hw2}
\end{equation}
where $f,f_z$ indicate the initial hyperfine state, while $s_1$, $s_2$, and $h_\nu$ denote the
spin $z$-projection for the two neutrons and the neutrino helicity state.
In turn, $T_{fi}(f,f_z;s_1,s_2, h_\nu)$ is given by
\begin{eqnarray}
T_{fi} (f,f_z;s_1,s_2,h_\nu) &\equiv&
\langle nn, s_1, s_2; \nu, h_\nu \,|\, H_W \,|\,
(\mu,d);f,f_z \rangle
\nonumber \\
&\simeq& {G_V' \over \sqrt{2}} \psi_{1s}M_{fi}(f,f_z;s_1,s_2,h_\nu)\,,
 \label{eq:h2ffz}
\end{eqnarray}
where we have defined
\begin{equation}
  \begin{aligned}
M_{fi}(f,f_z;s_1,s_2,h_\nu)&=\sum_{s_\mu s_d}
\langle {1 \over 2}s_{\mu}, 1 s_d | f f_z \rangle\,
l_\sigma(h_\nu,\,s_\mu)\\
&\times\langle \Psi_{{\bf p}, s_1 s_2}(nn) | j^{\sigma}({\bf q}) |
\Psi_{s_d}(d)\rangle \ .
\end{aligned}
\end{equation}
with $l_\sigma$ and $j^\sigma$ being the leptonic and
hadronic current densities respectively~\cite{Marcucci2011}.
Here the leptonic momentum transfer ${\bf q}$ is defined
as ${\bf q} \simeq -{\bf k}_\nu$ and ${\bf p}=({\bf p}_1-{\bf p}_2)/2$
is the relative momentum among the two neutrons.
Furthermore, $\Psi_d(s_d)$ and $\Psi_{{\bf p}, s_1 s_2}(nn)$ are the
deuteron and final $nn$ wave functions, respectively,
with $s_d$ indicating the deuteron spin $z$-projection,
which is computed using the variational
method described in Refs.~\cite{Marcucci2011,Ceccarelli2023}.
The final $nn$ wave function
$\Psi_{{\bf p}, s_1 s_2}(nn)$
can be expanded in partial waves as
\begin{eqnarray}
\Psi_{{\bf p},s_1 s_2}(nn)&=&4\pi\sum_{S} \langle \frac{1}{2} s_1,\frac{1}{2}s_2 |
S S_z \rangle 
\sum_{L L_z J J_z}{\rm i}^L Y^*_{LL_z}({\hat{\bf p}})\nonumber\\
&\times&
\langle S S_z, L L_z | J J_z\rangle \,\overline{\Psi}_{nn}^{LSJJ_z}(p) \>\> ,
\label{eq:psinnpw}
\end{eqnarray}
where $\overline{\Psi}_{nn}^{LSJJ_z}(p)$ is the $nn$ wave function with
orbital angular momentum $L L_z$, total spin $S S_z$, and total angular
momentum $J J_z$ that is computed numerically by using the
Kohn variational principle (see Ref.~\cite{Kohn1948}). The calculation is
performed using partial waves up to $J=4$. The contribution to the total capture rate of the partial waves with $J>2$ is of the order of $0.75\%$. We verified that  partial waves with $J>4$ give negligible contributions. 
Finally, in Eq.~(\ref{eq:h2ffz}),
the function $\psi_{1s}$
is the average over the nuclear volume of the muon
wave function in $1s$ orbit~\cite{Marcucci2011,Walecka1995}, namely
\begin{equation}
|\psi_{1s}|\simeq|\psi_{1s}^{\rm av}| \equiv\,  |\psi_{1s}(0)|\,=\,
\sqrt{{(\alpha\, \mu_{\mu d})^3\over \pi}} \ ,
\label{eq:psimud}
\end{equation}
where $\psi_{1s}(0)$ denotes the Bohr wave function
for a point charge $e$ evaluated at the origin, 
$\mu_{\mu d}$ is the reduced mass of the $(\mu,d)$ system,
and $\alpha$ is the fine-structure constant.
The integration of the matrix elements $M_{fi}$ is performed using
Gaussian-Legendre quadrature with $\approx45$  points
on the angles and $\approx 80$ on the inter-nucleon distance $r$.
This permits full convergence of the integrals.

Without losing generality we can choose ${\bf q}\parallel \hat{z}$
and define the angle $\theta_1$ as the angle between ${\bf q}$ and
${\bf p}_1$. After exploiting the conservation of energy in Eq.~(\ref{eq:fermigd})
the differential capture rate reads
\begin{eqnarray}\label{eq:diff_xsec}
  \Gamma^f(E_1')&=&\frac{G_V^{'2}}{\pi}|\psi_{1s}(0)|^2 E_1 p_1
  \int d\cos\theta_1 \frac{E_2 k_\nu^2}{E_2+k_\nu+p_1\cos\theta_1}\nonumber\\
  &\times&\sum_{s_1 s_2 h_\nu}\sum_{f_z}
  |M_{fi}(f,f_z,s_1,s_2,h_\nu;p_1,\cos\theta_1)|^2
\end{eqnarray}
where $k_\nu$ and $E_2$ can be easily obtained by the momentum and
energy conservation. The superscript $f$, that can be equal to 1/2 and 3/2, indicates the hyperfine state for which the capture rate is computed. 
Note that in this case the scattering
wave function depends explicitly on $p_1$ and $\cos \theta_1$
through ${\bf p}$ making the calculation much more expensive.

The total capture
rate is then computed integrating directly on the kinetic
energy $E_1$
\begin{equation}
  \Gamma^f=\int_0^{E_1^{'\rm max}}dE_1'\,\Gamma^f(E'_1)\,.
\end{equation}
The integrations
on $E_1'$ and $\cos \theta_1$ has been performed using Gauss-Legendre quadrature.
To reach convergence we need to use
at least 50 points on $E_1'$ and 20 on $\cos\theta_1$.
The total capture rate was also computed using the standard 
approach as in Ref.~\cite{Marcucci2011},
obtaining on the total capture rate numerical differences below 0.1 s$^{-1}$ for each nuclear interaction considered.
In Table~\ref{tab:cost} we report the constants and the masses
used in this work.  Note that the vector coupling constant that
first appears in Eq.~(\ref{eq:h2ffz}) is
\begin{equation}
  G_V^{'2}=G_V^2(1+\Delta_R^V)\,,
\end{equation}
where the vector coupling constant $G_V$ is given by $G_V=1.1357\times 10^{-11}$ MeV$^{-2}$, and the process independent radiative correction  $\Delta_R^V$ is
$\Delta_R^V=0.02454$, according to the new updated
values of Ref.~\cite{Hardy2020}. The final value
is then $G_V'=1.149\times 10^{-11}$ MeV$^{-2}$.

\begin{table}
  \centering
  \begin{tabular}{ll}
    \hline
    \hline
    $G_V'$      & $1.149\times 10^{-11}$ MeV$^{-2}$\\
    $1/\alpha$ & $137.04$ \\
    $m_d$      & $1875.61$ MeV \\
    $m_n$      & $939.57$ MeV \\
    $m_\mu$    & $105.66$ MeV \\
    \hline
    \hline
  \end{tabular}
  \caption{\label{tab:cost} Values of the
    constants used in the present calculation.}
\end{table}

\section{Nuclear interactions and electroweak currents}
\label{sec:models}
The interactions we use in the present calculation are of two types.
The first ones, developed in Norfolk (NV)~\cite{Piarulli2015,
  Piarulli2016},  are  local interactions 
up to the next-to-next-to-next-to-leading-order (N3LO), and include $\Delta$-isobars together with pions and
nucleons as degrees of freedom. The interactions are regularized in
configuration-space with two regulators, one ($R_S$) for
the short-range components associated with $2N$
contact terms, and the other ($R_L$) for the long-range terms.
We consider four different interactions of this family for which
two different sets of regulators have been used.
The LECs have been fitted  considering the $2N$ database within two different energy ranges.

The second family of interactions considered in this work is the one developed by
Entem, Machleidt and Nosyk (EMN) in Ref.~\cite{Entem2017}.
These interactions are implemented in momentum space and are strongly
non-local. The degrees of freedom are pions and nucleons only.
For this interaction family all the orders up to
the next-to-next-to-next-to-leading-order (N3LO)
are available for three different cutoff values
$\Lambda\,$=$\,450$, $500$ and $550$ MeV. The LECs of these
interactions are fixed fitting the $2N$ database up to $300$ MeV. 
In Table~\ref{tab:potlist} we  summarize the
names of the interactions used in this work
and their specific characteristics.
\begin{table}
  \centering
  \begin{tabular}{cccccc}
    \hline
    \hline
    Name & DOF & $O_\chi$ & $(R_{\rm S},R_{\rm L})$ or $\Lambda$ & $E$ range & Space \\
    \hline
    NVIa  & $\pi,N,\Delta$ & N3LO & $(0.8,1.2)$ fm & 0--125 MeV & $r$\\
    NVIb  & $\pi,N,\Delta$ & N3LO & $(0.7,1.0)$ fm & 0--125 MeV & $r$\\
    \hline
    NVIIa & $\pi,N,\Delta$ & N3LO & $(0.8,1.2)$ fm & 0--200 MeV & $r$\\
    NVIIb & $\pi,N,\Delta$ & N3LO & $(0.7,1.0)$ fm & 0--200 MeV & $r$\\
    \hline\hline
    EMN450     & $\pi,N$        & N3LO & $450$ MeV      & 0--300 MeV & $p$\\
    EMN500     & $\pi,N$        & N3LO & $500$ MeV      & 0--300 MeV & $p$\\
    EMN550     & $\pi,N$        & N3LO & $550$ MeV      & 0--300 MeV & $p$\\
    \hline
    \hline
  \end{tabular}
  \caption{\label{tab:potlist}Summary of $2N$ interactions used in this
study. In the first column we indicate the name adopted to identify each interaction
and in the remaining columns we list its main features, including degrees of freedom (DOF), chiral order ($O_\chi$),
cutoff values, lab-energy range over which the fits to the $2N$ database have been carried out ($E$ range), and
whether it is  expressed in configuration ($r$) or in momentum ($p$) space. }
\end{table}

The adopted models for the nuclear axial and vector
currents are the ones derived in Refs.~\cite{Baroni20162,Baroni2018} for the NV potentials
and Refs.~\cite{Pastore2009,Piarulli2013} for the EMN ones, respectively. In this work we performed the analysis on the truncation errors of the currents considering both the Bochum (see for example Ref.~\cite{Krebs2020}) and the JLab-Pisa group power counting (see for example Refs.~\cite{Pastore2009,Piarulli2013}). We summarize the various contributions to the currents for the Bochum and JLab-Pisa power counting in Tables~\ref{tab:currents_bochum} and~\ref{tab:currents}, respectively. Note that for the Bochum power counting we considered the relativistic corrections of the same order as in the naive-power counting.\footnote{This has been done to maintain consistency with the LECs fitted in previous works. Clearly this generates some theoretical inconsistency. However, from the numerical point of view the total contribution of the relativistic corrections are of the order of $\approx1\%$ on the total capture rate, and the impact is minimal on the error analysis. For completeness we report in Table~\ref{tab:currents} the N4LO contribution of the currents for the JLab-Pisa group power counting.}  

%
\begin{table}
  \centering
  \begin{tabular}{lcccc}
    \hline
    \hline
    Oper. & LO $(Q^{-3})$ & NLO $(Q^{-2})$ & N2LO $(Q^{-1})$ & N3LO $(Q^{0})$ \\
    \hline
    \multirow{2}{*}{$\rho(A)$} & \multirow{2}{*}{--} &
      \multirow{2}{*}{--} & 1b(NR) & \multirow{2}{*}{--} \\
      &   &  & OPE  &  \\
      & & & & \\
      \multirow{2}{*}{${\bf j} (A)$} &\multirow{2}{*}{1b(NR)}&
      \multirow{2}{*}{--} & OPE-$\Delta^*$ & CT($d_R$) \\
      &   &  & [1b(RC)]  &  OPE \\  
      & & & & \\
      $\rho(V)$ & 1b(NR) & -- & [1b(RC)] & [OPE(RC)] \\ 
      & & & & \\
       \multirow{2}{*}{${\bf j}(V)$} & \multirow{2}{*}{--} & \multirow{2}{*}{--}& 1b(NR) &
      OPE-$\Delta^*$\\
      & & & OPE & [1b(RC)] \\
      \hline
    \hline
  \end{tabular}
  \caption{\label{tab:currents_bochum}
Ordering of the chiral electroweak
    currents as given in Ref.~\cite{Krebs2020}. The acronym stands for 1b=one-body, OPE=one-pion exchange, CT=contact terms, NR=non-relativistic,
    RC=relativistic corrections, and  OPE-$\Delta$ = one-pion-exchange currents with an intermediate $\Delta$-isobar excitation. The terms in the square bracket are RC that we kept at the 
    order given by the naive power-counting.
    With the star  we indicate the terms that do not appear for the
    EMN interactions.}
\end{table}
\begin{table*}
  \centering
  \begin{tabular}{lccccc}
    \hline
    \hline
    Oper. & LO $(Q^{-3})$ & NLO $(Q^{-2})$ & N2LO $(Q^{-1})$ & N3LO $(Q^{0})$ & N4LO $(Q^{1})$ \\
    \hline
    \multirow{2}{*}{$\rho(A)$} & \multirow{2}{*}{--} &
      \multirow{2}{*}{1b(NR)} & \multirow{2}{*}{OPE} & \multirow{2}{*}{--} & TPE$^A$ \\
      &   &  &  &  & CT$^\dag$ \\
      \hline
      \multirow{2}{*}{${\bf j} (A)$} &\multirow{2}{*}{1b(NR)}&
      \multirow{2}{*}{--} & 1b(RC) & CT($d_R$) & TPE\\
      &   &  & OPE-$\Delta^*$ &  OPE  & OPE(sub) \\
      \hline
      $\rho(V)$ & 1b(NR) & -- & 1b(RC) & OPE(RC) & TPE \\ 
      \hline
      \multirow{3}{*}{${\bf j}(V)$} & \multirow{3}{*}{--} &
      \multirow{3}{*}{1b(NR)}& \multirow{3}{*}{OPE} & 1b(RC) &
      TPE \\
      & & & & OPE-$\Delta^*$ & OPE($d_2^V$, $d_3^V$, $d_2^S$) \\
      & & & &                & CT($d_1^V$, $d_1^S$) \\
      
      \hline
    \hline
  \end{tabular}
  \caption{\label{tab:currents} The same as Table~\ref{tab:currents_bochum} for the JLab-Pisa group power counting scheme of the electroweak currents. For completeness we report also the contributions at N4LO derived by our group. The acronym stands for 1b=one-body, OPE=one-pion exchange, CT=contact terms, TPE=two-pion exchange, NR=non-relativistic, RC=relativistic corrections, OPE-$\Delta$ = one-pion-exchange currents with an intermediate $\Delta$-isobar excitation, and sub=sub-leading. With the asterisk we indicate the terms that do not appear for the EMN interactions. The term with the superscript A are not yet available for the NV interactions. The dagger indicates that the LECs appearing in these terms have not been determined yet. Note that we explicitly show for each term the LECs that have been fitted on electroweak processes.}
  \end{table*}

One of the goal of this work is to study the role of the axial contact term (CT) at N3LO and therefore of the  value of the LEC $d_R$ (see Eq.~(\ref{eq:A1})) on the determination of the total capture rate. 
Since the LEC $d_R$ is linearly dependent on the LEC $c_D$ that appears in the three-nucleon interaction,
this has been determined fitting contemporary the $^3$H
binding energy and the Gamow-Teller matrix element of the $^3$H
$\beta$-decay. 

For the NV interactions we used the value of $c_D$ (and $c_E$) fitted in  Ref.~\cite{Baroni2018}.
For the EMN interactions we refitted $c_D$ (and $c_E$) following the procedure of Ref.~\cite{Baroni2018}. The results are reported in Appendix~\ref{app:cd}.


The calculation of the differential and total
muon capture rate on deuteron presented below has been carried out
for all the nuclear interactions presented
in Table~\ref{tab:potlist} and for all the chiral order from
LO to N3LO in the
case of the EMN interactions.

\section{Analysis of the theoretical uncertainties}
\label{sec:errors}

In this section we focus on
the sources of uncertainties and on how we dealt with them.
The sources of uncertainties that we consider are four:
(i) the uncertainties on the LECs appearing in the nuclear electroweak currents as they result from the fitting procedure, and on the single-nucleon axial form factor;
(ii) the error due to the truncation of the chiral expansion of the current;
(iii) the error due to the truncation of the chiral expansion of the interaction;
(iv) the dependence on the nuclear interaction model.
Clearly, also the LECs fitted in the nuclear interaction have an impact
on the determination of the full uncertainties as well.
However, a comparison of the results obtained 
with different nuclear interactions as in point (iv)
can partially give an estimate of this impact.
We are going to consider point (iv) in Section~\ref{sec:discussion} where
we will combine the results obtained in for the single interactions.
Note also that the four sources of uncertainty are not necessarily
independent. However in this work we treat them as if they were.
Moreover, the power counting used for organizing the current terms plays a crucial role in the final determination of the uncertainty. While this cannot be considered as a source of error it-self, we computed the uncertainties on the currents considering both available power countings in order to give the most comprehensive  picture of the present theoretical situation.

In Table~\ref{tab:results} we present the computed total capture
rate for the hyperfine state 1/2 for the various nuclear interactions considered together with
the error associated with the various sources of uncertainty considering the Bochum and JLab-Pisa group power counting respectively. In the Table
we report the results of the EMN considering the interaction at N3LO.
Since the N4LO currents are not fully determined, we performed our analysis considering the currents only up to N3LO using the value of $d_R$ fitted consistently at N3LO as well. We discuss in detail in the following subsections all the various sources of uncertainties.
In Table~\ref{tab:results} we present also the results (without errors) for the capture rate in
the hyperfine state 3/2. These are consistent with the results of Ref.~\cite{Bonilla2023} once the extra contribution of the partial waves $J=3$ and $J=4$ is removed ($\approx0.75\%$).

\begin{table*}
  \centering
  \begin{tabular}{lllcccccc}
    \hline
    \hline
    Name & pot. & curr. & $\Gamma^{1/2}$ &
    $\sigma^C_{k=3}$[BPPC]  & $\sigma^C_{k=3} $[JPPC] 
   &  $\sigma^I_{k=4}$ &    
    $\sigma_{\rm LECs}$ & $\Gamma^{3/2}$ \\
    \hline
    NVIa   & N3LO & N3LO & 393.5 & 5.1$^*$(3.0) &  1.1(0.7) & n.a.& 3.9 & 13.3\\
    NVIb   & N3LO & N3LO & 393.7 & 5.1(3.0)& 1.1(0.7) & n.a.& 3.9 & 13.5\\
    NVIIa  & N3LO & N3LO & 392.5 & 5.1(3.0)& 1.1$^*$(0.7) & n.a.& 3.9 & 13.3 \\
    NVIIb  & N3LO & N3LO & 392.6 & 5.1(3.0)& 1.1(0.7) & n.a.& 3.9 & 13.3\\
   & & & & & & \\                                                                
    EMN450 & N3LO & N3LO & 396.0 & 6.4(3.1)& 2.2(0.7) & 0.3$^*$(0.2) & 3.9 & 13.7 \\
    EMN500 & N3LO & N3LO & 397.3 & 6.0(3.1)&  2.2$^*$(0.7) & 0.3$^*$(0.2) & 3.9 & 13.8 \\
    EMN550 & N3LO & N3LO & 397.0 & 6.2(3.1)&  2.1$^*$(0.7)& 0.4$^*$(0.2) & 3.9 & 13.8 \\
    \hline                                 
    \hline                                 
  \end{tabular}                            
  \caption{\label{tab:results} Total muon capture rate on deuteron for all the interactions considered in this work. With $\Gamma^{1/2(3/2)}$($J\leq4$) we indicate the computed value using the currents and the interactions at N3LO with partial waves up to $J=4$ in the singlet (triplet) hyperfine state. $\sigma^{C(I)}_{k=3(4)}$ 
  is the standard deviation of the truncation
  error computed using the Bayesian framework of Ref.~\cite{Melendez2019} for the current (interaction). For the truncation error associated to the current we report the values obtained using both the Bochum group (BPC) and the JLab-Pisa group power counting (JPPC). The values
  between parenthesis in the fifth, sixth and seventh columns are the ones obtained using the prescription
  of Ref.~\cite{Epelbaum2015}. Finally in the eighth column we report the error computed propagating the 
  uncertainties associated with the LECs appearing in the currents. All the uncertainties are reported at $68\%$ confidence level (CL). The star on the truncation error indicates that the emulator failed the statistic tests for some specific cases (see text for more details).} 
\end{table*}                               
                                           
\subsection{Current LECs uncertainties}    
\label{sec:LECs_error}

The axial nuclear charge and current operators are multiplied by the
single-nucleon axial form factor  $g_A(q^2)$, with $q$ indicating
the four-momentum transfer. The single-nucleon
axial form factor can be parametrized as
\begin{equation}
  g_A(q^2)=g_A\left(1-\frac{1}{6}r_A^2 q^2\right)\,,
\end{equation}
where $g_A=1.2723$~\cite{Patrignani2016} for the NV interactions and $g_A=1.2754$~\cite{PDG} for the EMN interactions, and we adopted $r_A^2 = 0.46(16)$ fm$^2$,
as suggested in Ref.~\cite{Hill2018}.
Since $q^2$ for the muon capture on deuteron  is quite large, the uncertainty on $r_A^2$ makes a significant
 impact on the capture rate.
At the same time is important to study the impact of the uncertainty on the LEC $d_R$ on the total capture rate. Such uncertainty is given in Refs.~\cite{Baroni2018} for the NV interactions and Appendix~\ref{app:cd} for EMN interactions.
The errors on $r_A^2$ and $d_R$ have been propagated with standard error propagation techniques, i.e.
\begin{equation}\label{eq:slec}
  \sigma^2_{\rm LECs}=\left(\frac{\partial \Gamma}
    {\partial r_A^2}\right)^2\sigma^2(r_A^2)+
    \left(\frac{\partial \Gamma}
         {\partial d_R}\right)^2\sigma^2(d_R)\,.
\end{equation}
The $68\%$ confidence level (CL) results can be found
in Table~\ref{tab:results}. The computed uncertainties
are identical within
the showed digits for all the nuclear interactions considered. The reason is that the values of the derivatives in Eq.~(\ref{eq:slec}) are constant and almost independent of the interactions ($\partial \Gamma/\partial r_A^2=-24$ s$^{-1}$ fm$^{-2}$, and $\partial \Gamma/\partial d_R=0.66$ s$^{-1}$).
The uncertainty we obtain is slightly smaller respect to Ref.~\cite{Bonilla2023} (4.4 s$^{-1}$). This is due to the fact the single-nucleon axial form factor is associated to each term in our axial current and charge operators, while in Ref.~\cite{Bonilla2023} it appears only at LO. This slightly reduce the absolute value of the derivative respect to $r_A^2$ in Eq.~(\ref{eq:slec}) and therefore the error associated with it. For example, removing the axial form factor from the higher order terms in the current using the NVIa interaction, we obtain an error of 4.3 s$^{-1}$  consistent with Ref.~\cite{Bonilla2023}.

The impact of the $d_R$ error on $\sigma_{\rm LECs}$ 
is of the order of $1\%$, and therefore completely negligible.
This is a consequence of the small contribution that the contact
term of the axial current at N3LO gives to the total muon capture.
We also tested the impact of the errors on the LECs appearing at N4LO in the vector part of the current (see Table~\ref{tab:currents}) obtaining results similar to $d_R$.

\subsection{Bayesian analysis of truncation error} 
The analysis of the uncertainties due to the truncation of the chiral expansion
for currents and interactions
is performed using the gsum package\footnote{Some of the libraries where slightly modified for
addressing the $p$ dependence when computing the truncation error.}
within the formalism introduced by Melendez {\it et al.} in Ref.~\cite{Melendez2019}. We first review the fundamental points of the Bayesian analysis
needed to present our specific case. 

\subsubsection{Brief introduction}\label{sec:trunc_intro}
We introduce here the main concepts used in our Bayesian analysis of the truncation error in $\chi$EFT. We will refer to Ref.~\cite{Melendez2019}
for all the remaining theoretical and technical details
behind the analysis. Note that we assume that
the chiral expansions of the currents and the interactions
are independent. Therefore,
we are going to study them separately, keeping
fix the interaction order 
at N3LO when we study the truncation error of the current expansion.
In order to study the truncation error of the chiral interaction expansion,
we keep the order of the chiral current fixed at N3LO
when the interaction order is N2LO and N3LO,
and at N2LO when the interaction
is used at LO and NLO.
This choice is made because $d_R$ is not defined for the interaction
at LO and NLO. 
We are going to indicate with a superscript $C$($I$) the
specific quantities that are relative to
the analysis of the truncation error of the
current (interaction). The equation where these
indexes are missing are valid for both the cases.

The observable we consider for this analysis is the differential radiative capture $\Gamma(E_1')$ defined in Eq.~(\ref{eq:diff_xsec}) which
depends on the kinetic energy of the neutron $E_1'$. This
is directly connected with
the neutron momentum $p_1=\sqrt{{E_1'}^{2}+2E_1'm_n}$,
which we are using as independent variable (see Ref.~\cite{Melendez2019}).
For notation clarity in the
next two subsections we are going to write the quantity
$\Gamma$ as function of the momentum of the neutron $p_1$ only.

The $k$-th order EFT prediction can be written as
\begin{equation}
  \Gamma_k(p_1)=\Gamma_{\rm ref}(p_1)\sum_{n=0}^{k}c_n(p_1)Q^n(p_1)\,,
\end{equation}
where $\Gamma_{\rm ref}(p_1)$ is a dimension-full overall scale that
is selected such that the dimensionless coefficients $c_n$
are of order 1. 
As in Ref.~\cite{Melendez2019} we take
\begin{equation}\label{eq:QQ}
  Q(p_1)=\frac{1}{\Lambda_b}\frac{p_1^{8}+m_\pi^8}{p_1^{7}+m_\pi^7}\,,
\end{equation}
with $m_\pi$ the mass of the pion, and $\Lambda_b$ the breakdown scale
of the theory. In this work we follow Refs.~\cite{Melendez2019,Acharya2022} taking $\Lambda_b=600$ MeV, which is a reasonable value between the cutoffs of the interactions and the
formal breaking scale energy of chiral effective field theory (i.e. 1 GeV) .
The EFT truncation error is then defined as
\begin{equation}
  \delta\Gamma_k(p_1)=\Gamma_{\rm ref}(p_1)\sum_{n=k+1}^{\infty}c_n(p_1)Q^n(p_1)\,.
\end{equation}
Our goal is then to determine this truncation error and the uncertainty associated
with it.

The idea of Ref.~\cite{Melendez2019} is to build a stochastic representation of the $c_n(p_1)$
based on a Gaussian Process (GP) that emulates our order-by-order chiral calculation.
The GP is then exploited to emulate the missing $c_n$ that appears in the EFT truncation error.
The basic assumption is that $c_n(p_1)$ are identical independent draws of an underling GP, i.e.
\begin{equation}\label{eq:assumption}
  c_n(p_1) \mid \mu, \bar{c}^2, \ell \stackrel{\text { iid }}{\approx} \mathcal{G P}\left[\mu, \bar{c}^2 r\left(p_1, \overline p_1 ; \ell\right)\right]\,,
\end{equation}
where this is specified by the mean $\mu$, the variance $\bar{c}^2$ and the correlation length $\ell$.
The correlation function $r\left(p_1, \overline p_1 ; \ell\right)$ is assumed to have an exponential-squared form (see Ref.~\cite{Melendez2019}).
The GP hyper-parameters $\mu$, $\bar{c}^2$, and $\ell$ are then learned from the training data set
which contains order-by-order EFT calculations. The truncation error distribution is then given by~\cite{Melendez2019}
\begin{equation}\label{eq:gkp}
\delta \Gamma_k(p_1) \mid \mu, \bar{c}^2, \ell, \Lambda_b \approx \mathcal{G} \mathcal{P}\left[m_{k}(p_1), \bar{c}^2 R_{k}\left(p_1,\overline  p_1 ; \ell\right)\right]\,,
\end{equation}
where
\begin{equation}
m_{k}(p_1)=\Gamma_{\mathrm{ref}}(p_1) \frac{Q(p_1)^{k+1}}{1-Q(p_1)} \mu\,,
\end{equation}
and
\begin{equation}
  \begin{aligned}
  R_{k}\left(p_1, \overline p_1 ; \ell\right)&=\Gamma_{\mathrm{ref}}(p_1) \Gamma_{\mathrm{ref}}
  \left(\overline p_1\right)\\
  &\times \frac{\left[Q(p_1) Q\left(\overline p_1\right)\right]^{k+1}}{1-Q(p_1) Q\left(\overline p_1\right)} \,r\left(p_1, \overline p_1 ; \ell\right)\,.
  \end{aligned}
\end{equation}
Note that in Eq.~(\ref{eq:gkp}) we assume that the truncation error distribution depends
only on $\Lambda_b$ and not on the choice of the functional 
form of $Q(p_1)$ which we assume fixed.
The distribution for the full observable will have then mean and covariance respectively given by
\begin{equation}
    \Gamma_{\rm th}(p_1)=\Gamma_k(p_1)+m_{k}(p_1)\,,
\end{equation}
and
\begin{equation}
    \Sigma_{\rm th}(p_1,\overline p_1,\ell)=\bar{c}^2 R_{k}\left(p_1,\overline p_1 ; \ell\right)\,.
\end{equation}

Finally, the total capture rate is the integral of
$\Gamma(p_1)$ on $p_1$ (this can be directly derived
from Eq.~(\ref{eq:diff_xsec}) after changing variables).
To all the practical purposes this integral can be
written as a discrete sum over $M$ points, i.e
\begin{equation}\label{eq:int_th} \Gamma_{\rm th}=\sum_{i=1,M}\omega_i\;\Gamma_{\rm th}(p_1(i))\,,
\end{equation}
where $\omega_i$ are the specific weights of the chosen
integration method. From Eq.~(\ref{eq:int_th})
it follows immediately
that $\Gamma_{\rm th}=\Gamma_k+\delta\Gamma_k$, with
$\Gamma_k(\delta\Gamma_k)$
the integration over $p_1$ of $\Gamma_k(p_1)(\delta\Gamma_k(p_1))$. 
What we want is then
to find the distribution of
\begin{equation}
  \delta \Gamma_k=\sum_{i=1,M}\omega_i\, \delta \Gamma_k(p_1(i))\,.
\end{equation}
Using the properties of Gaussian random variables we find
\begin{equation}\label{eq:GKP}
  \delta \Gamma_k \mid \mu, \bar{c}^2, \ell, \Lambda_b 
  \approx \mathcal{N}\left[M_{k}(\mu), \sigma^2_{k}(\bar{c}^2, \ell, \Lambda_b)\right]\,,
\end{equation}
where $\mathcal{N}$ indicates the  normal distribution, and
\begin{equation}
  M_{k}{(\mu)}=\sum_{i=1,M}\omega_i\, m_{k}(p_1(i))\,,
\end{equation}
and
\begin{equation}\label{eq:tot_err}
  \sigma^2_{k}(\bar{c}^2,\ell,\Lambda_b)=\bar{c}^2\sum_{i,j=1,M}\omega_i\omega_j\, R_{k}\left(p_1(i), p_1(j) ; \ell\right)\,.
\end{equation}
Note that Eq.~(\ref{eq:GKP}) is not anymore a Gaussian process since
we now have a single random variable.
With this final equation we can discuss the specific details of our analysis. 

\subsubsection{Currents truncation errors}\label{sec:trunc_cur}
For the analysis of the truncation error of the chiral expansion of the currents we consider the factorization
\begin{equation}
  \Gamma^C_k(p_1)=\Gamma^C_{\rm ref}(p_1)\sum_{n}c^C_n(p_1)Q^n(p_1)\,,
\end{equation}
where $Q^n(p_1)$ has been defined in Eq.~(\ref{eq:QQ}), $n=\{0,2,3\}$ for the Bochum group power counting and $n=\{0,1,2,3\}$ for the JLab-Pisa one,
and we select the reference scale as
\begin{equation}
  \Gamma^C_{\rm ref}(p_1)=\frac{\Gamma^C_{\rm LO}(p_1)Q^{-3}(p_1)}{10}\,.
\end{equation}
In such a way we reconstruct the correct power of $Q(p_1)$,
such that the calculation does not suffer of inversion
problems, and the $c^C_n(p_1)$ are naturally sized (this is the reason for the factor $1/10$). We remove from the analysis the coefficient $c^C_0(p_1)$ because the choice of the reference value makes it not significant for training the emulator.

We decide to limit our analysis up to $p_1^{\rm max}=195$ MeV.
This permits us to verify the hypothesis that the $c_n^C(p_1)$ are
distributed as in Eq.~(\ref{eq:gkp}) using the diagnostics presented in Ref.~\cite{Melendez2019}. Beyond  $p_1^{ \rm max}$ the coefficients $c_n^C(p_1)$ become very large and  the hypothesis of Eq.~(\ref{eq:gkp}) results not statistically valid anymore.
The most reasonable explanation is the fact that the scale $Q$ in Eq.~(\ref{eq:QQ}) is not the proper one for this process at such large $p_1$. Note that the contribution to the total capture rate of the tail beyond $p_1^{\rm max}$ is $\lesssim 2$ s$^{-1}$. Therefore, the impact on the error computation is minimal.

The data set generated using the Bochum group power counting consists of 195 data points on a grid
that starts from zero and has a step of 1 MeV.
 For training
the emulator we use 4(5) data points distant 50(40) MeV for the
NV(EMN) interactions. We used more points in the EMN case because of the larger oscillations of the coefficients as function of $p_1$. As validation set we used 13
 data points distant 15 MeV one from each other.
More data points in the validation data set give rise to ill-defined covariance matrices.
Finally, we use $10^{-4}$ as the
value for the variance of the white noise needed to stabilize the
matrix inversion (i.e. the nugget). Smaller values
generate instabilities in the inversion of the covariance matrices,
while values
larger than $10^{-3}$ generate too much noise in the final results.

\begin{figure*}[bth]
  \centering
  \subfigure[]{\includegraphics[scale=0.7]{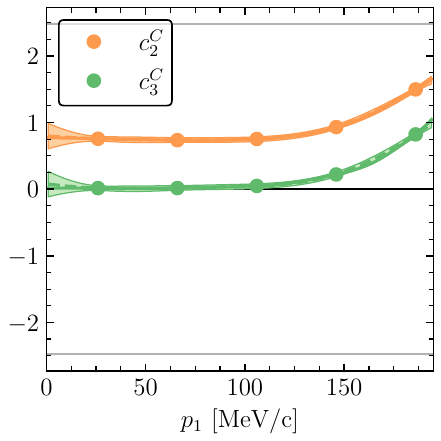}}
  \subfigure[]{\includegraphics[scale=0.7]{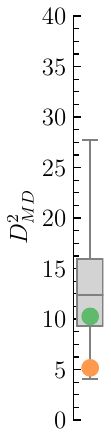}}
  \subfigure[]{\includegraphics[scale=0.7]{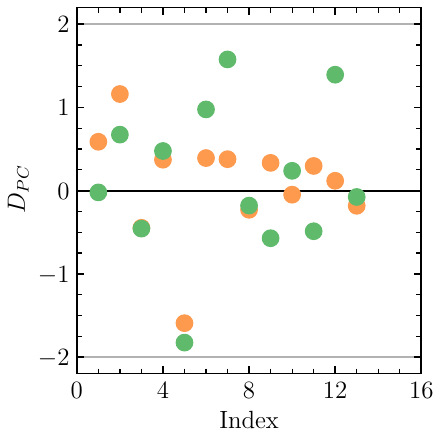}}
  \subfigure[]{\includegraphics[scale=0.7]{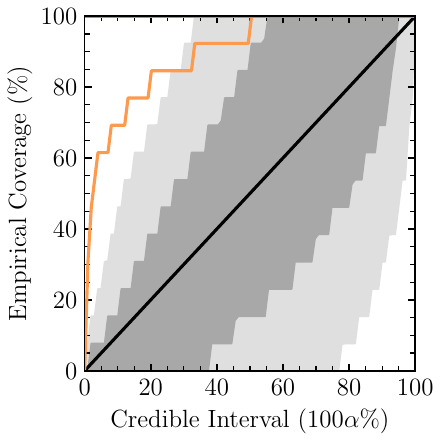}}
  \caption{\label{fig:trunc_curr}The Gaussian Process modeling of
    the current chiral expansion coefficients and its diagnostics for
    the EMN550 interaction using the Bochum power counting. In figure (a) the simulators
    (solid lines - i.e. our calculation) along with the corresponding
    Gaussian process emulators (dashed lines) and their $2\sigma$ intervals
    (bands). The data used for training are denoted by dots.
    (b) The Mahalanobis distances compared to the mean (interior line),
    $50\%$ (box) and $95\%$ (whiskers) credible intervals of the
    reference distribution. 
    (c) The pivoted Cholesky diagnostics versus the index along with $95\%$
    credible intervals (gray lines). 
    (d) The credible interval diagnostics for the truncation error bands.
    The  $1(2)\sigma$ is represented with the dark(light) gray band. The steepness of the orange line indicates that the N2LO and the N3LO are very close to each other (see text for more details). }
\end{figure*}
\begin{figure*}[bth]
  \centering
  \subfigure[]{\includegraphics[scale=0.7]{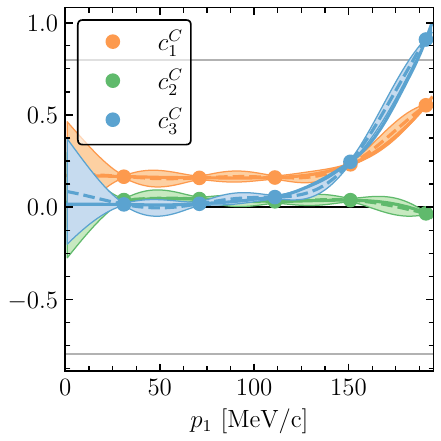}}
  \subfigure[]{\includegraphics[scale=0.7]{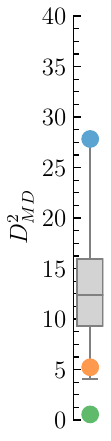}}
  \subfigure[]{\includegraphics[scale=0.7]{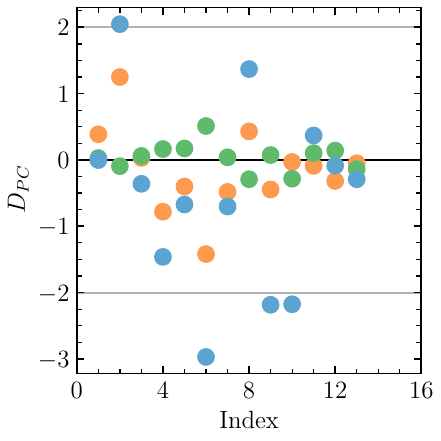}}
  \subfigure[]{\includegraphics[scale=0.7]{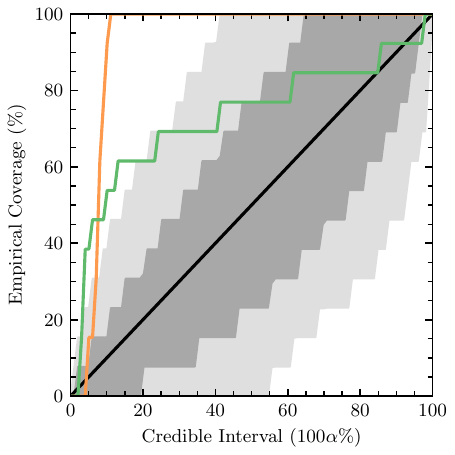}}
  \caption{\label{fig:trunc_curr_JP}The The same as Figure~\ref{fig:trunc_curr} for
    the EMN550 interaction using the JLab-Pisa power counting. }
\end{figure*}
In Figure~\ref{fig:trunc_curr}(a) we report the value of the coefficients $c_n^C(p_1)$
together with the GP emulator results and their $2\sigma$ interval for the EMN550 interaction using the Bochum power counting. 
From a first inspection, the GP emulator is
able to nicely reproduce the $\chi$EFT calculation.
To asses quantitatively
the quality of the emulation we performed the Mahalanobis distance ($D^2_{MD}$) test
presented in Figure~\ref{fig:trunc_curr}(b),
and the pivoted Cholesky decomposition ($D_{PC}$) test presented in
Figure~\ref{fig:trunc_curr}(c).
The Mahalanobis distance test is a generalization of the squared residuals in the
case of correlated data points (see Ref.~\cite{Melendez2019} for more details).
The comparison with the reference
$\chi^2$ distribution shows a compatibility of the emulation with
the data points 
within the $95\%$ CL (whiskers) for both $c^C_2$
and $c^C_3$. 
This is evident also at the level of the more informative Cholesky
decomposition (see Ref.~\cite{Melendez2019}).
The validation points are distributed almost uniformly within 1$\sigma$, with only few points close to the $2\sigma$ line.
Similar results on the two diagnostics are obtained for all the other interactions, except for the NVIa model. In this case, the casual cancellation of the current contributions at N3LO give rise to coefficients $c_3^C$ practically zero. Therefore, almost no stochastic fluctuations are generated  in the emulator. This gives rise to an anomalous coincidence among the emulator and the simulator results, making the emulator fail the Mahalanobis and the pivoted Cholesky test for the NVIa interaction (i.e. the simulator seems to work statistically too good). The final error is in any case in reasonable agreement with the others and we report in Table~\ref{tab:final_err} with a star.

\begin{figure}[bth]
 {\includegraphics[width=1\columnwidth]{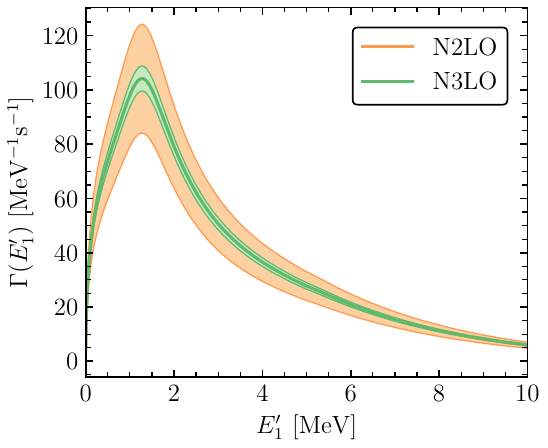}}
  \caption{\label{fig:spectra_curr}The differential capture
    rate as function of the neutron energy $E_1'$ computed with the EMN550
    interaction fixed at N3LO for various order of the
    current N2LO (orange), and N3LO (green) using the Bochum power counting. The bands represent the $2\sigma$ truncation errors at each order.}
\end{figure}

In the credible interval diagnostic showed 
in Figure~\ref{fig:trunc_curr}(d) we study
if the truncation error computed 
at each order is compatible with the correction at the next order
within a certain CL.
The CL bands are constructed by sampling
a large number of emulators (1000) from the underlying process.
 The credible intervals are then plotted
 against the percentage of validation points found within the interval,
 i.e. if the emulator contains only a small
 amount of the validation points is over-confident (in the figure represented as horizontal lines),
 in the other case is under-confident (vertical lines).
The orange line in Figure~\ref{fig:trunc_curr}(d)
is the credible interval for $\Delta \Gamma_2^C(p_1)=\Gamma^C_2(p_1)-\Gamma^C_3(p_1)$
compared with the reference distribution of $\Delta \Gamma^C_2(p_1)$, which is also the only one we can compute performing this analysis.
 The credible interval for $\Delta \Gamma^C_2(p_1)$ shows us that the emulation is in general under-confident. That is understandable since the differences among the simulated spectra at N2LO and N3LO is minimal
 (see Figure~\ref{fig:spectra_curr}).
 The results for the confidence interval diagnostics of the other interactions are practically identical.

  \begin{figure}[bth]
 {\includegraphics[width=1\columnwidth]{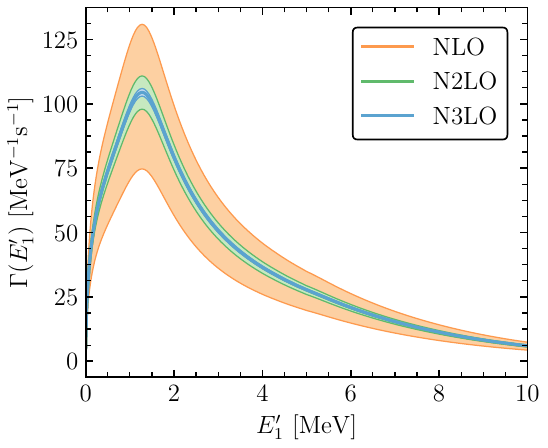}}
  \caption{\label{fig:spectra_curr_JP}The differential capture
    rate as function of the neutron energy $E_1'$ computed with the EMN550
    interaction fixed at N3LO for various order of the
    current NLO(orange), N2LO (green), and N3LO (blue) using the JLab-Pisa power counting. The bands represent the $2\sigma$ truncation errors at each order.}
\end{figure}
 A similar analysis has been performed considering the data generated using the JLab-Pisa power counting for the NV(EMN) interaction up to 195(190) MeV with 1 MeV step. For training the emulator we use 5 data points distant
35(40) MeV. As validation set we used 13 data points distant 15 MeV one from each other. The nugget we use is $10^{-4}(2\times 10^{-4})$. 

The results for the diagnostic of the EMN550 interaction are shown in Figure~\ref{fig:trunc_curr_JP}. Again the emulator seems able to reproduce nicely the coefficients $c_n^C(p_1)$. However, the emulator starts to have difficulties for $p_1>150$ MeV, since the coefficients $c_n(p_1)$ become very stiff, especially $c^C_3$ (see Figure~\ref{fig:trunc_curr_JP}(a)). Consequently, the Mahalanobis distance test is slightly failed by the emulator for the coefficient $c_3^C$ (Figure~\ref{fig:trunc_curr_JP}(b)). 
This is confirmed also by the pivoted Cholesky test where several blu points are out of the $2\sigma$ range (Figure~\ref{fig:trunc_curr_JP}(c)). Note also the statistically anomalous coincidence of the emulator with the simulator for the coefficients $c_1^C$. The analysis of the EMN500 interaction give almost identical results, while for the EMN450 the emulator pass all the tests including the  $c_3^C$ coefficient. For the NV interaction the emulator pass all the tests again except in the case of the $c_3^C$ coefficient for the NVIIa interaction that present rather strong oscillation that the emulator is not able to address completely.

In Figure~\ref{fig:trunc_curr_JP}(d)
 we show the credible interval for $\Delta \Gamma_1^C(p_1)=\Gamma^C_1(p_1)-\Gamma^C_2(p_1)$ (green), and $\Delta \Gamma_2^C(p_1)=\Gamma^C_2(p_1)-\Gamma^C_3(p_1)$ (orange)
compared with the reference distribution. The vertical behavior for both cases indicate that the emulator is under-confident, i.e. the error estimated is statistically larger than expected. This indicate also that the error generated at given order contain the result at next order. Even if the emulator is not able to describe all the features of the coefficients $c_n^C$, the truncation error estimate is reliable even if rather conservative.
 
 The diagnostic gave us an overall good result on the
 quality of the GP emulation. We computed then the truncation error using Eq.~(\ref{eq:gkp}).
 The order-by-order spectra obtained from the EMN550
 interaction with $95\%$ CL truncation error for both the Bochum and JLab-Pisa power counting are shown in Figure~\ref{fig:spectra_curr} and Figure~\ref{fig:spectra_curr_JP} respectively.
 Finally, we compute the truncation error on the total capture rate as in
 Eq.~(\ref{eq:tot_err}) for each interaction. The $68\%$ CL
 results are reported in Table~\ref{tab:results}.
 Note that the integration  has been performed only up to $p_1^{\rm max}=195$ MeV.
 In order to estimate the error arising from the remaining part of the
 spectra we used the Epelbaum {\it et al.} prescription (EP)~\cite{Epelbaum2015}
 as considered in Ref.~\cite{Ceccarelli2023}.
 By doing so, we have found that the contribution to the total truncation 
 error of this part of the spectra is completely negligible. Similar results 
 have been obtained for all the other interactions.

 In Table~\ref{tab:results} we report between parentheses also the errors
 computed on the total capture rate using the EP exactly as in Ref.~\cite{Ceccarelli2023}.
 To give a more statistical insight,  we assume that the expected decay rate is
 uniformly distributed within the limits settled by the extreme values and
 so the $68\%$ CL is given by the value of the truncation error divided
 by $\sqrt{3}$. As it can be seen from the table, in general the error estimated with
 the EP is smaller than the one obtained using
 the Bayesian analysis.

\subsubsection{Interaction truncation error{\sout{s}}}\label{sec:trunc_int}

The other source of uncertainties we treat with the Bayesian analysis
is the one arising from the truncation of the nuclear interaction chiral expansion. Unfortunately,
we do not have all the orders of the NV interactions. Therefore, we limit the
analysis only to the EMN interaction family.
The procedure that we use is identical to the one used in the analysis of the current
truncation errors. The main difference
is in the  factorization we use, chosen to be
\begin{equation}
  \Gamma^I_k(p_1)=\Gamma^I_{\rm ref}(p_1)\sum_{n=0,2,3,4}c^I_n(p_1)Q^n(p_1)\,,
\end{equation}
where $Q^n(p_1)$ has been defined in Eq.~(\ref{eq:QQ}) and
$\Gamma^I_{\rm ref}(p_1)=\Gamma^I_{\rm LO}(p_1)$.
This choice of the reference value makes the coefficient
$c^I_0$ useless for  training the emulator since it results identical to one.
Therefore we exclude it in our analysis. A similar choice was done in Refs.~\cite{Melendez2019,Acharya2022}.

To obtain a reasonable result for the GP emulator we restrict the momentum range of the
analysis to  180 MeV. For training the emulator we use 8 points distant
25 MeV and  we take 21 test points 8 MeV apart from each other. The nugget
we used in this case  is $2\times 10^{-4}$. Note that
we need a higher density of training points compared to the currents because of the large
oscillations of the coefficients $c^I_n$.

\begin{figure*}[bth]
  \centering
  \subfigure[]{\includegraphics[scale=0.7]{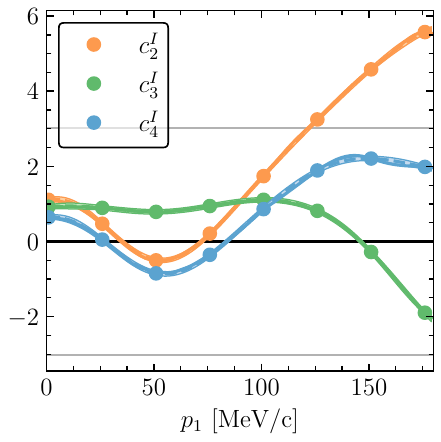}}
  \subfigure[]{\includegraphics[scale=0.7]{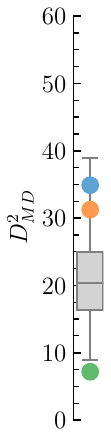}}
  \subfigure[]{\includegraphics[scale=0.7]{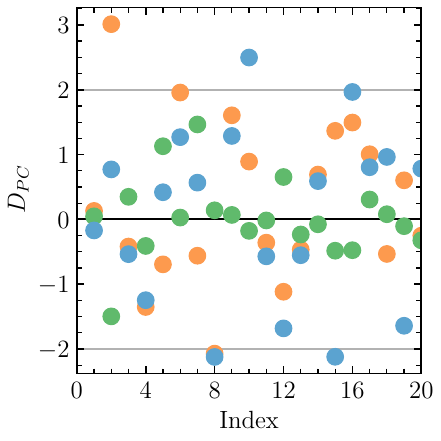}}
  \subfigure[]{\includegraphics[scale=0.7]{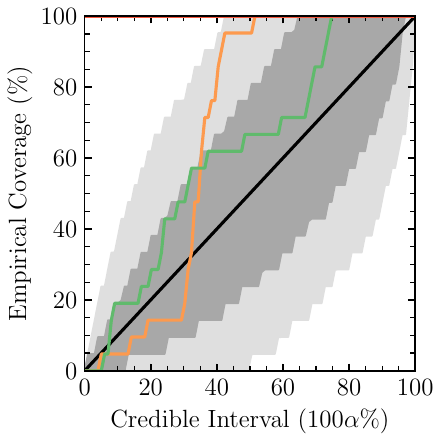}}
  \caption{\label{fig:trunc_int}Same as Figure~\ref{fig:trunc_curr}, but for the truncation errors of the interaction chiral expansion. The results reported in the figure are obtained using the EMN550 interaction fixing the order of the chiral current at N2LO for the interaction at NLO, and at N3LO in the remaining cases. }
\end{figure*}
In Figures~\ref{fig:trunc_int} we present the results of the emulation together with
the diagnostic performed to verify the quality of the emulation trained on the EMN550 interaction model results.
As it can be observed from the figure, the Mahalanobis distance (Fig.~\ref{fig:trunc_int}(a)) the coefficient $c_3^I$ show a slightly anomalous coincidence among  the simulator and the emulator, confirmed by a high density of green points close to zero in the 
Cholesky decomposition (Fig.~\ref{fig:trunc_int}(b)). A similar behaviour is present for the coefficient $c_4^I$ in the case of the EMN500 interaction. For the EMN450 the emulator fails the Mahalanobis distance test for the $c_4^I$ as well. In this case the reason is the large oscillation of this coefficient at large momentum $p_1$ that makes the emulator very hard to train.

In order to check if we could improve the emulation, we performed several other  tests changing the parameters of the analyses without finding any significant improvement on the statistical tests. Despite this, the truncation errors obtained with these other analysis are numerically identical to the one obtained using the parameters shown in the text.

On the other hand, the credible interval diagnostic test (Figure~\ref{fig:trunc_int}) for all the EMN interactions
a compatibility of more than $95\%$ between the truncation error at given
order and the prediction at the next one up to N3LO over all the empirical coverage.
This indicates that at each order the estimated error contains the next order correction.   
Therefore, despite the emulator is not able to pass the statistical tests for each $c_n^I$, the estimate of the truncation error seems to be reliable. 

 \begin{figure}[bth]
  \includegraphics[width=1\columnwidth]{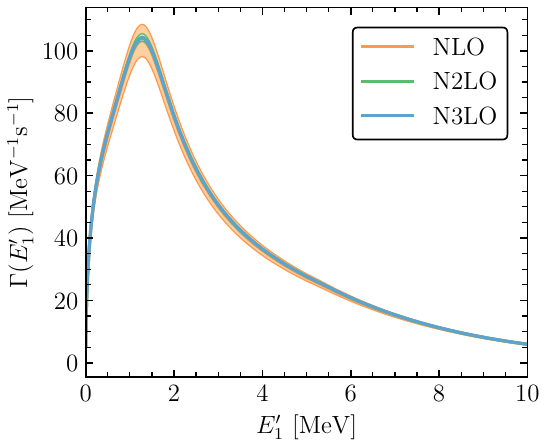}
  \caption{\label{fig:spectra_pot}The differential capture
    rate as function of the neutron energy $E_1'$ computed with the EMN550
    interaction at NLO (orange), N2LO (green) and N3LO (blue). The bands
    represent the $2\sigma$ truncation errors at each order.
    Note that the chiral order of the current used in this 
    analysis is N2LO when the interaction is at NLO and
    N3LO in the rest of the cases.}
\end{figure}
We present in Figure~\ref{fig:spectra_pot} the differential
 capture rate spectrum computed order-by-order with the relative
 truncation error for the EMN550 interaction. The truncation error on the
 total capture rate obtained using Eq.~(\ref{eq:tot_err}) has been reported
 in Table~\ref{tab:results}. Once again we have verified, using the EP, that the error
 arising from the tail of the spectra not analyzed in the Bayesian procedure is negligible. 
 The error obtained applying the EP over all the spectrum can be found in Table~\ref{tab:results} between parentheses.

 \subsubsection{Discussion}
Before concluding this section some remarks are in order.
\begin{itemize}
\item The power counting of the currents has a major impact on the determination of the truncation errors. As can be seen from our results the Bochum group power counting gives  larger error bands than the JLab-Pisa. This has been seen already in Ref.~\cite{Martin2023} for the electromagnetic terms of the current. 

\item Even if the results for all the interactions are compatible within the error bars, the capture rates obtained using the local interactions are systematically lower than the ones obtained using the non-local interactions. This is partially due to the different sign of the contact terms in the axial current at N3LO.

\item In this work we chose to study separately the truncation error associated to the interaction and the currents. Clearly this is not  completely correct because of precise relations between the Hamiltonian and the currents such as current conservation. Even if such relations exist, they are not completely fulfilled order-by-order by the available currents yet.  We did some attempt to compute the truncation errors for the Hamiltonian and the currents together. However, we did not find any benefit for the analysis, obtaining on the other side an error typically smaller than the current truncation error reported here. 
Therefore we decided to consider here the most conservative approach.

\end{itemize}

\section{Final results and discussion}
\label{sec:discussion}

We consider now the last source of uncertainty, i.e.\ the one arising from model dependence. Compared with previous studies, where essentially a reference value was obtained as the mean over all the interactions considered and the error was estimated as the difference between the extreme cases, we follow a slightly different approach as proposed in Ref.~\cite{William2021} and often used in Lattice QCD for model averaging.

We take for each of the model as our best estimate of the total capture rate the one computed, i.e. we assume
\begin{equation}\label{eq:tot_mean}
  \Gamma_i(\infty)=\Gamma_i({\rm comp})\,,
\end{equation}
where $i$ indicate the model used in the calculation.
We consider the worst scenario in which all the three sources of error are fully correlated and so the total error is given by
\begin{equation}  \sigma_i=\sigma^C_{i,k=3}+\sigma^I_{i,k=4}+\sigma_{i,{\rm LECs}}\,.
\end{equation}
Note that as $\sigma^C_{i,k=3}$ we consider only the one obtained using the Bochum power counting that represent the worst case scenario. For the NV interaction family we take $\sigma^I_{k=4}=0.4$ s$^{-1}$
as in the worst case of the EMN interactions.

Following Ref.~\cite{William2021} the average of the models is given by
\begin{equation}
\langle \Gamma \rangle =\sum_i \Gamma_i\; {\rm pr}(i)
\end{equation}
while the variance can be written as
\begin{equation}
\sigma^2_{\Gamma}=\sum_i \sigma^2_i\; {\rm pr}(i)+\sigma^2_{\Gamma,{\rm syst}}\,,
\end{equation}
with the systematic error given by the model dependence obtained as 
\begin{equation}
\sigma^2_{\Gamma,{\rm syst}}=\sum_i \Gamma_i^2 \; {\rm pr}(i)-\left(\sum_i \Gamma_i\; {\rm pr}(i)\right)^2\,.
\end{equation}
In these equations, with "${\rm pr}$" we have indicated the probability of a certain model.
Since there is no reason to privilege local or non-local interactions, and no reason to privilege any interaction within a given class, we assign the following probabilities
\begin{equation}
    {\rm pr}(i)=\begin{cases} \frac{1}{8}, & \mbox{if } i\mbox{ local } \\ \frac{1}{6}, & \mbox{if } i\mbox{ non-local} \end{cases}.
\end{equation}

Using the Bochum group power counting we obtain 
\begin{equation}
  \Gamma_{\rm th}({\rm BPC})=(395\pm 10)\,{\rm s}^{-1}\quad (68\% {\rm CL})\,,
\end{equation}
while using the JLab-Pisa power counting
\begin{equation}
  \Gamma_{\rm th}({\rm JPPC})=(395\pm 6)\,{\rm s}^{-1}\quad (68\% {\rm CL})\,.
\end{equation}
 As recommended value we select the more conservative result obtained using the Bochum group power counting.

\begin{figure}[bth]
  \includegraphics[width=1\columnwidth]{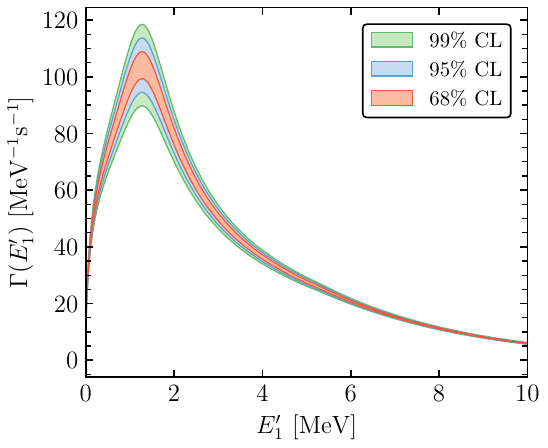}
  \caption{\label{fig:spectra_final} The recommended differential capture
    rate as function of the kinetic neutron energy $E_1'$. The green, blue, and red
    bands represent respectively the 99$\%$, 95$\%$, and $68\%$ CL. Note that the difference among the bands can be appreciated
    only at the peak of the spectra.}
\end{figure}

An identical analysis have been performed on the differential capture rate for each value of $E_1'$.
Our recommended spectra obtained using the Bochum group power counting  is shown in Figure~\ref{fig:spectra_final} with the bands at $68\%$, $95\%$, and $99\%$ CLs.  A table with the numerical values of these final spectra is provided as supplementary material \cite{suppl}. 

\begin{figure}[bth]
  \includegraphics[width=1\columnwidth]{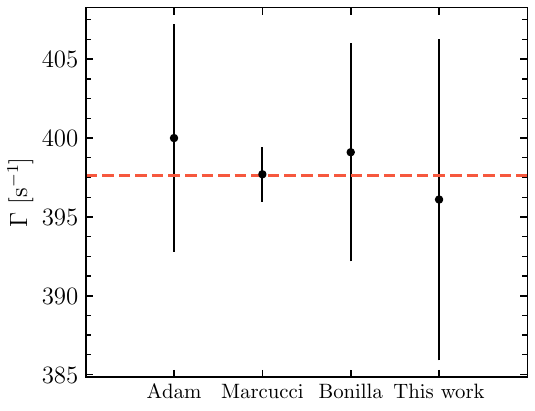}
  \caption{\label{fig:comparison} 
  Comparison of the theoretical results
    obtained in our work using the Bochum (BPC) and JLab-Pisa (JPPC) power counting with the  $\chi$EFT results of Refs.~\cite{Adam2011,Marcucci2012chiral,Bonilla2023}.
    The errors in the previous works have been assumed to be the limit of a
    uniform distribution and then divided by $\sqrt{3}$ to obtain the $68\%$
    CL. Note that only in Ref.~\cite{Bonilla2023} and in the present
    work various sources of uncertainty besides the model
    dependence were considered. The red dashed line represent the arithmetic mean of all the theoretical results.  }
\end{figure}

The comparison of our results with those present in the literature still obtained within $\chi$EFT
can be seen in Figure~\ref{fig:comparison}\footnote{ For the Marcucci {\it et al.}~\cite{Marcucci2012chiral} we report the result in the Erratum.}. 
For the
calculations in Refs.~\cite{Marcucci2012chiral,Bonilla2023} we have assumed that the errors represent the
extreme of a uniform distribution. For Ref.~\cite{Adam2011}, we have considered as limit of the distribution the minimal and the maximal reported results.
Therefore, in Figure~\ref{fig:comparison} we report the
error divided for $\sqrt{3}$, in order to obtain the $68\%$ CL. Note also that 
only in our work and in Ref.~\cite{Bonilla2023} the error 
includes the uncertainty on $r_A^2$.

The capture rates obtained in this work using the local interactions are systematically  smaller than the ones obtained using the non-local ones. Therefore, the overall result is slightly smaller compared with the world literature in which only non local chiral interactions have been used. On the other hand all the theoretical $\chi$EFT calculations are consistent  within $1\sigma$.
We do not proceed in comparing our results with the available experimental data of Refs.~\cite{Wang1965,Bertin1973,Bardin1986,Cargnelli1989}, because of their large uncertainties.

In order to appreciate the importance of the various uncertainty sources in the total error budget, we list in Table~\ref{tab:final_err} their absolute and
relative weight in percentage for the two considered power counting.
In the JLab-Pisa power counting the main source of uncertainty results $r_A^2$ while in the case of the Bochum power counting the current truncation error become the dominant source.
 The model dependence slightly increases the total error
but it is smaller than the error generated by the current truncation error and $r_A^2$. 
 Note that the truncation error obtained here using the Bochum power counting is similar to
truncation error estimated in Ref.~\cite{Bonilla2023}, using
the prescription of Ref.~\cite{Furnstahl2015}, once properly rescaled for
the different value of $\Lambda_b$ used in our work.
\begin{table}
  \centering
  \begin{tabular}{lcc}
    \hline
    \hline
    Uncertainty source & BPC & JPPC \\
    \hline  
     $r_A^2$ & $5.6(30.8\%)$ & $5.4(75.2\%)$ \\
     Other current LECs & \multicolumn{2}{c}{negligible} \\
    $\chi$EFT truncation - currents & $8.1(65.2\%)$ & $2.4(14.6\%)$\\
    $\chi$EFT truncation - interactions & $0.5(0.3\%)$ & $0.5(0.7\%)$ \\
    Model dependence & $1.9(3.7\%)$ & $1.9(9.5\%)$\\
    \hline
    \hline
  \end{tabular}
  \caption{\label{tab:final_err} Absolute and relative contributions to
  the estimated total theoretical error from the various uncertainty
  sources for both the Bochum (BPC) and JLab-Pisa (JPPC) power counting.}
\end{table}

We want to underline that reducing the uncertainty on $r_A^2$ is crucial for testing pure $\chi$EFT effects on the muon capture on deuteron at few percent level. In this sense recent Lattice QCD results on the computation of the
nucleon axial form factors are extremely encouraging (see Ref.~\cite{Meyer2022} for a review). However, while experimental efforts are on going to reduce the experimental error, improvements on the theoretical side for reducing the truncation errors are crucial to extract physical parameters from this observable.

\section{Impact on the MUSUN experiment}\label{sec:musun}
In this section we present a minimal study of the impact
of our results in the analysis of the future experimental results of MuSun. For our analysis we assume that the final error of the experiment is the expected 
precision of $\approx1.5\%$~\cite{Kammel2021}. Since this is just
a preliminary analysis we consider only the nuclear interaction NVIa and the error associated to this interaction reported in the first line of Table~\ref{tab:results}. For this analysis we assume that the MuSun experiment finds a central value for the total capture rate of $393.5$ s$^{-1}$ and an error of $5.9$ s$^{-1}$ ($1.5\%$).

First we study the dependence of the total capture rate as
function of the value of the LEC $c_D$ from which $d_R$ depends linearly.  This is mostly known as

\begin{figure}[bth]
  \includegraphics[width=1\columnwidth]{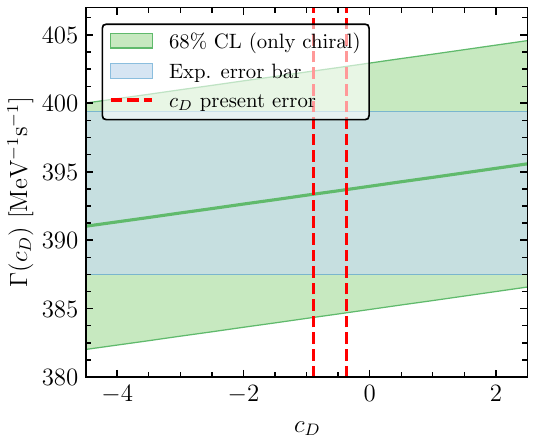}
  \caption{\label{fig:C_D_dependance} Capture rate as function
  of the value of $c_D$ for the NVIa interaction. The green band represent the error on $\Gamma^{3/2}$ due to the truncation error on the chiral expansion of the current and the interaction. The red dashed lines represent the present error bar on $c_D$. The blue horizontal line is the error band corresponding to an hypothetical result of MuSun $\Gamma^{3/2}=393.5$ s$^{-1}$ with a $1.5\%$ total error. }
\end{figure}

As it can be seen in Figure~\ref{fig:C_D_dependance} the dependence of $\Gamma$ on $c_D$ is linear but with a very small slope. Indeed, just considering only the chiral truncation errors (green band), the experimental value with an uncertainty of $\approx 1.5\%$ (blue horizontal band) would not have any impact on the improvement of the present $c_D$ constraints (red vertical dashed lines). Indeed a $1.5\%$ precision on the experimental result corresponds to a very large set of values of $c_D$, much larger than the present constrains. 

\begin{figure}[bth]
  \includegraphics[width=1\columnwidth]{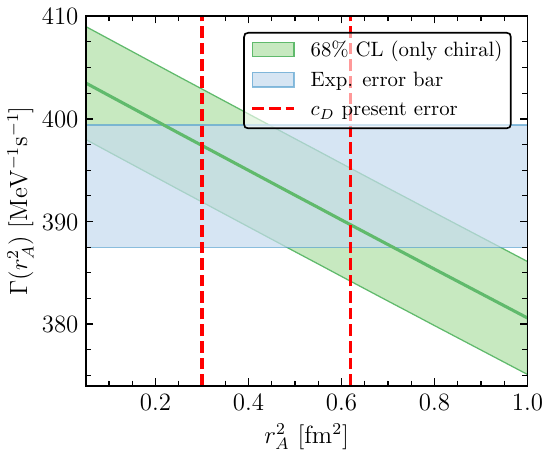}
  \caption{\label{fig:r_A_dependance} Capture rate as function
  of the value of $r_A^2$ for the NVIa interaction. The green band represent the error on $\Gamma$ due to the truncation error on the chiral expansion of the current and the interaction. The red dashed lines represent the $68\%$ CL on $\Gamma$ considering the error on $r_A^2$ as reported in Ref.~\cite{Hill2018}. The blue dashed lines represent the predicted $68\%$ CL on $\Gamma$ with an error of $10\%$ on $r_A^2$. }
\end{figure}

We can perform a similar analysis varying $r_A^2$. In this case for simplicity we keep $c_D$ fixed. In Figure ~\ref{fig:r_A_dependance} we plot the the capture rate as function of $r_A^2$ for the NVIa interaction. The red dashed lines represents the present uncertainty on $r_A^2$. The blue band again is the hypothetical result of MuSun with a $\approx 1.5\%$ error. As can be seen, the overlap region of the blue and green bands spans a region of values of $r_A^2$ much larger than the present limits on it. This indicates that considering our most conservative estimate  on the theoretical errors, the MuSun experiment with the present precision goal can not be a useful source for having an independent estimate of $r_A^2$. 

Although the present theoretical error bands does not allow, with the expected precision  of the MuSun experiment, to obtain new determination of fundamental constants, the MuSun experiment is still a fundamental test for the parameters of chiral effective field theory. 
Indeed, MuSun will be the first precise measurement of the rate for a weak process in the two-nucleon system, which can be compared with theoretical predictions accompanied by fully quantified uncertainties.  Any strong deviation of the experimental results or inconsistency with the present literature would imply the necessity of a profound revision for the chiral electroweak currents.
Furthermore, we would like to stress that the MuSun experiment remains one of the best options to access with good accuracy the LEC $L_{1A}$ present (see for example the work of Chen {\it et al.} \cite{Chen2005}), within pionless EFT, in several two-nucleon processes,
among which, besides muon capture, also the proton-proton fusion reaction,
of paramount importance in astrophysics.

\section*{Acknowledgments}
The Authors would like to thank Prof. P. Kammel for critical reading of the manuscript and precious suggestions.
A.G. would like to thank Prof. D. Phillips for useful discussion on the Bayesian analysis and the use of the gsum
package and A. Rodas for the suggestion on the model averaging. We thank the BUQEYE collaboration for making the gsum package available. The calculation was performed using resources of the National Energy Research Scientific Computing Center (NERSC), a U.S. Department of Energy Office of Science User Facility located at Lawrence Berkeley National Laboratory, operated under Contract No. DE-AC02-05CH11231.

\appendix
\section{Fit of the $c_D$ low energy constant for the EMN interactions}\label{app:cd}

We updated the fit of the values of the LECs $c_D$ and $c_E$ using the updated value for $g_A=1.2754$~\cite{PDG} and the tritium beta-decay Gamow-Teller matrix element GT=$0.9501\pm0.0024$  as in Fit-3 of Ref.~\cite{Acharya2023} and also~\cite{Baroni2016}. Note that we use as definition for $d_R$
\begin{equation}\label{eq:A1}
d_R=-\frac{M_N}{4 \Lambda_\chi g_A} c_D+\frac{1}{3} M_N\left(c_3+2 c_4\right)+\frac{1}{6}\,,
\end{equation}
where $M_N$ is the nucleon mass, $g_A$ the nucleon axial coupling  and $c_3$, $c_4$ are LECs with the values listed in the Table.
We refitted the two LECs using the procedure of Ref.~\cite{Marcucci2012chiral}, i.e.\ fixing the LECs $c_D$ and $c_E$ to reproduce  the $A = 3$ binding energies and the
Gamow-Teller (GT) matrix element of tritium $\beta$-decay. 
In Table~\ref{tab:tab_gt2} we present the results of the fit together with the values of the LECs $c_{1,3,4}$ used in the currents and the interactions.

\begin{table}[h]
    \centering
    \begin{tabular}{lcccccc}
    \hline
    \hline
         & $\Lambda$ & $c_1 $& $c_3$ & $c_4$ & $c_D$ & $c_E$ \\
         \hline
         N2LO & 450 & $-0.74$ & $-3.61$ & $2.44$ & $-0.13(20)$ & $-0.09(4)$ \\
         N2LO & 500 & $-0.74$ & $-3.61$ & $2.44$ & $-0.86(19)$ & $-0.32(4)$ \\
         N2LO & 550 & $-0.74$ & $-3.61$ & $2.44$ & $-1.93(20)$ & $-0.73(4)$ \\
              &     &     &   &  & & \\
         N3LO & 450 & $-1.07$ &  $-5.32$ & $3.56$ & $-0.42(20)$ & $\m0.06(5)$ \\
         N3LO & 500 & $-1.07$ & $-5.32$ & $3.56$ & $-2.73(23)$ & $-1.03(5)$ \\
         N3LO & 550 & $-1.07$ & $-5.32$ & $3.56$ & $-4.35(24)$ & $-2.16(5)$\\
              &     &     &   &  & & \\
         N4LO & 450 & $-1.10$ & $-5.54$ & $4.17$ & $\m0.17(20)$ & $\m0.04(5)$\\
         N4LO & 500 & $-1.10$ & $-5.54$ & $4.17$ & $-3.02(25)$ & $-1.21(4)$\\
         N4LO & 550 & $-1.10$ & $-5.54$ & $4.17$ & $-4.48(24)$ & $-2.35(4)$ \\
         \hline
         \hline
    \end{tabular}
    \caption{Values for the LECs $c_{1,3,4}$ , $c_D$, and $c_E$ at the chiral orders N2LO, N3LO, and N4LO. The $c_D$ and $c_E$ LECs reproduce the $A = 3$ binding energies and the GT matrix element in tritium $\beta$-decay, as explained in the text.}
    \label{tab:tab_gt2}
\end{table}
                         
\bibliography{bibliography.bib}
\end{document}